\documentclass{iau} 
\usepackage{graphicx}

\title[{[}WN{]} central stars of planetary nebulae] 
{[WN] central stars of planetary nebulae}

\author[Todt et al.]   
{H.~Todt$^1$,
B.~Miszalski$^2$,
J.~A.~Toal\'a$^{3,4}$,
\and M.~A.~Guerrero$^4$}

\affiliation{
$^1$Institute of Physics and Astronomy, 
    University of Potsdam, \\ 
    Karl-Liebknecht-Str.~24/25, 
    14476 Potsdam, Germany
    \\ email: {\tt htodt@astro.physik.uni-potsdam.de} 
    \\[\affilskip]
$^2$South African Astronomical Observatory,\\
PO Box 9, Observatory 7935, South Africa;\\
Southern African Large Telescope Foundation,\\
PO Box 9, Observatory 7935, South Africa
\\[\affilskip]
$^3$Institute of Astronomy and Astrophysics, Academia Sinica (ASIAA),\\
10617 Taipei, Taiwan, Republic of China 
\\[\affilskip]
$^4$Instituto de Astrof\'isica de Andaluc\'ia, IAA-CSIC,\\
Glorieta de la Astronom\'ia s/n, E-18008 Granada, Spain 
}

\pubyear{2017}
\volume{323}  
\jname{Planetary nebulae: Multi-wavelength probes of stellar and galactic evolution}
\editors{X. Liu, L. Stanghellini \& A. Karakas, eds.}
\begin{document}

\maketitle

\begin{abstract}
While most of the low-mass stars stay hydrogen-rich on their surface throughout
their evolution, a considerable fraction of white dwarfs as well as  central
stars of planetary nebulae have a hydrogen-deficient surface composition. 
The majority of these H-deficient central stars exhibit spectra very similar
to massive Wolf-Rayet stars of the carbon sequence, i.e. with broad
emission lines of carbon, helium, and oxygen. In analogy to the
massive Wolf-Rayet stars, they are classified as [WC] stars. Their
formation, which is relatively well understood, is thought to be the
result of a (very) late thermal pulse of the helium burning shell.  

It is therefore surprising that some H-deficient central stars which have
been found recently, e.g.\ IC\,4663 and Abell 48, 
exhibit spectra that resemble those of the massive
Wolf-Rayet stars of the nitrogen sequence, i.e. with strong emission
lines of nitrogen instead of carbon. This new type of central stars is
therefore labelled [WN]. We present spectral analyses of these
objects and discuss the status of further candidates 
as well as the evolutionary status and origin of the [WN] stars. 
\end{abstract}

\firstsection 
\section{Introduction}

Wolf-Rayet central stars are hydrogen deficient central stars of
planetary nebulae which exhibit in their spectra strong
emission lines of helium, carbon and
oxygen. Because their spectra resemble those of massive WC stars, they are
called [WC] stars, with brackets to distinguish
them from their massive counterparts. 
In spite of spectral similarities and comparable chemical
composition, the formation of the low-mass
[WC] stars is completely different from the formation of the massive WC
stars. 
Stellar evolutionary models accounting for simultaneous burning and
mixing explain the formation of a [WC] star by the occurrence of a
thermal pulse (TP) at the very end or after the asymptotic giant
branch (AGB) phase of a
H-normal low-mass star. These models predict a hydrogen-deficient
surface composition with carbon enriched up to $X_{\rm C}=40\%$
after a late or very late TP (\cite[Althaus et
  al.\ 2005]{althaus2005}, \cite[Herwig 2001]{herwig2001}). 
Only in the case of a very late TP (VLTP) a supersolar nitrogen abundance of
about $X_{\rm N}=1\%$ is expected, but without any remaining hydrogen. 
Thus low-mass central stars with WN-like surface abundances are theoretically
not expected.

\section{The [WN] stars} 

\subsection{PB\,8}
The CS PB\,8 was classified  as [WC\,5-6] by \cite[Acker \& Neiner (2003)]{acker2003} 
based on a low-resolution spectrum. 
A spectral analysis based on a high resolution spectrum, 
\cite[Todt et al.\ (2010)]{todt2010} revealed that this object has a stellar temperature
of about $50\,$kK and an unusual
composition with He:H:C:N:O=55:40:1:1:1 by mass and resembles
spectroscopically a massive WN/C star.

Due to its unknown distance, also $L$ is unknown, so whether
this star is a true [WN/C] CS can only be inferred indirectly.

The nebula analysis by \cite[Garc\'ia-Rojas (2009)]{garcia-rojas2009} yield values for
$T_{\rm e}$ and $n_{\rm e}$ which are typical for young  PNe, as well as
 the  small $v_{\rm exp}\approx 30\,$km/s.
Regarding the  luminosity distance, one finds that
for a CSPN with $L/L_\odot=6000$, PB\,8 would be at a
distance of 4.2\,kpc and at a height of 300\,pc above the Galactic
plane, whereas a massive WR star would have at least
$L/L_\odot=2\times10^5$, and so at least a distance of 24\,kpc and a height of
1.7\,kpc above the Galactic plane. This is a rather untypical location
for a massive WR star.
So, PB\,8 is indeed a CSPN.

We also considered the possibility that PB\,8 may be a binary.
Although this can not be excluded completely, it is rather unlikely as there
were no shifts of radial velocities of spectral
lines detected \cite[(M\'endez 1989)]{mendez1989} and the 
nebula appears spherically symmetric, also in velocity space.

\subsection{IC\,4663}

The first ``pure'' [WN] was discovered by 
\cite[Miszalski et al.\ (2012)]{miszalski2012}. 
Their spectral analysis based on an optical spectrum 
revealed that IC\,4663 is
an almost hydrogen-free ($<2\,\%$ by mass) [WN] star, 
whose wind consists to 95\,\% of helium with only 0.8\% nitrogen.
Interestingly it is of early spectral subtype ([WN3]) with a temperature
of about 140\,kK and a relatively low mass-loss rate of about 
$2\times10^{-8}\,M_\odot\,{\rm a}^{-1}$. \cite[Miszalski et
  al.\ (2012)]{miszalski2012} showed that if
IC~4663 were a massive WN star it would be at an implausible distance 
of 58\,kpc and more than 8\,kpc below the Galactic plane. Moreover, they
discovered an AGB halo around IC~4663, proving it is a CSPN.

\subsection{Abell\,48}

\begin{figure}[h]
\begin{center}
 \includegraphics[width=\textwidth]{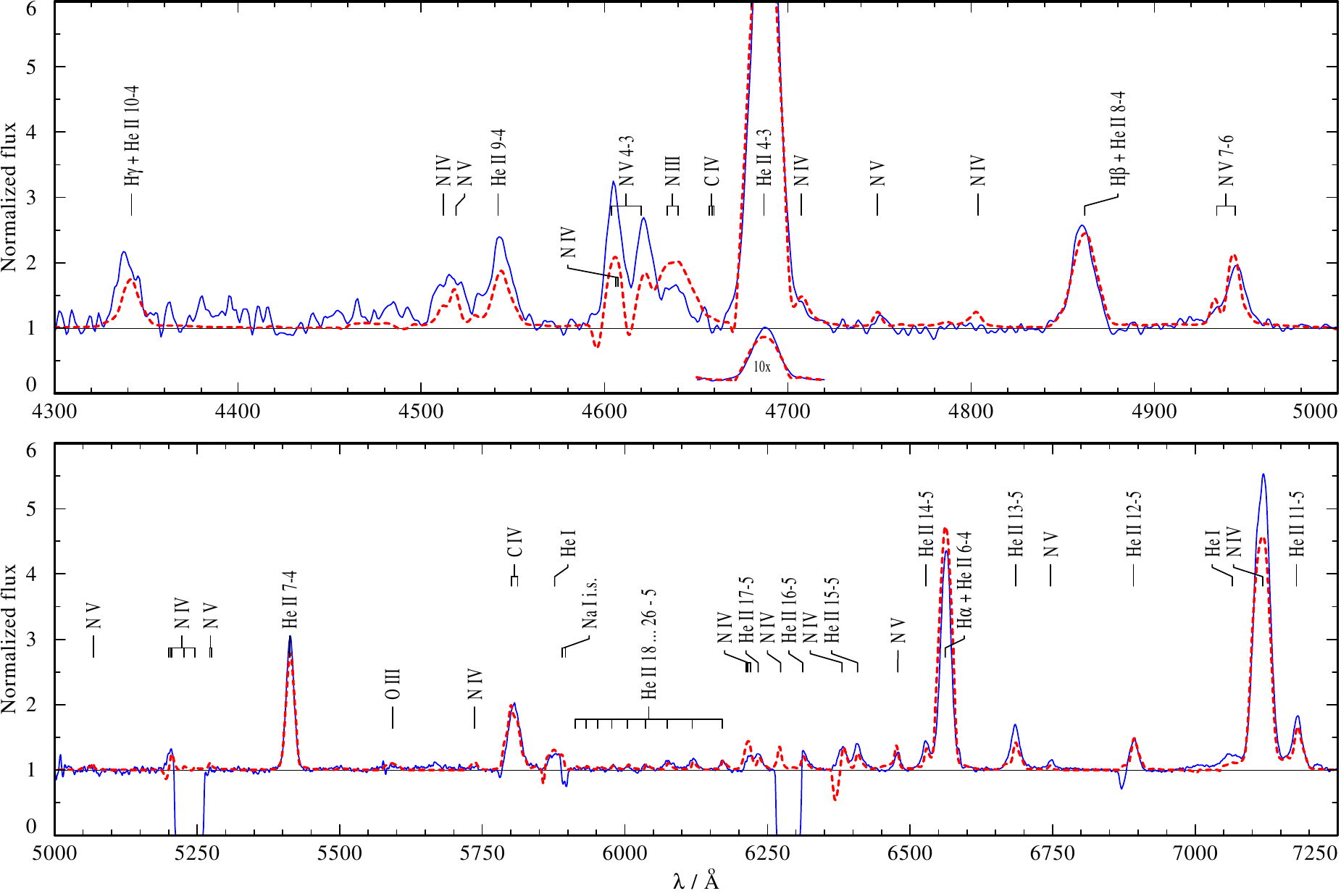} 
 \caption{Optical spectrum of Abell\,48: observation (blue, thin line)
 and best-fitting PoWR model (red, thick dotted line; see \cite[Todt
   et al.\ 2013]{todt2013}), both normalized to the model continuum}
   \label{fig:abell48}
\end{center}
\end{figure}

\cite[DePew et al. (2011)]{depew2011} mentioned the CS of Abell\,48,
but it was unclear whether this is really a CSPN and not a massive WN
star (\cite[Wachter et al. 2010]{wachter2010} classified it as WN6).
The CS Abell\,48 was found to have a helium-rich (85\%) wind with 
about 10\% hydrogen, and a nitrogen abundance of about $5\%$
(\cite[Todt et al.\ 2013]{todt2013}, see Fig,~\ref{fig:abell48}; \cite[Frew et
  al.\ 2014]{frew2014}). 
From the nebula analysis 
(line ratios of S\,{\sc ii} and H$\alpha$, 
as in \cite[Riesgo-Torado \& L\'opez 2002]{riesgo2002})
it was concluded that Abell\,48 is indeed a CSPN. This is also consistent with
the measured extinction and its significant proper motion - Abell\,48 is a runaway. 
The CS has a mass-loss rate of $4\times 10^{-7}\,M_\odot\,{\rm
  a}^{-1}$ and a temperature of about 70\,kK. Its spectral subtype is [WN5].

\section{[WN] candidates}

The earliest claim for the discovery of a [WN] central star was by 
\cite[Morgan et al. (2003)]{morgan2003}, who observed the
highly reddened ($E_{{\rm B}-{\rm V}} = 3\,{\rm mag}$) object
PMR\,5, which shows a WN spectrum (subtype WN6) and a round nebula.
However, it is not clear, whether this is a PN or ring nebula around a
massive WR star. While the electron density is consistent with a PN, 
the expansion velocity of 
$v_{\rm exp}=165\,$km/s is rather unusual for a PN, but typical for a
ring nebula.
Moreover, the luminosity distance is consistent with
a  massive WR star.
  For a typical CSPN luminosity of $L/L_\odot=6000$ PMR\,5 would be
  located at a distance of 0.5\,kpc and at a height of 6\,pc above the
  Galactic plane, while for a typical luminosity of a massive WN star
  PMR\,5 would be at 2.9\,kpc away and 35\,pc above the Galactic
  plane. Hence, both locations are consistent with the corresponding type 
of star. However, a reddening of $E_{{\rm B}-{\rm V}}=3\,$mag is rather 
untypical for  a distance of only 500\,pc. 
A spectral analysis based on the published spectrum revealed a chemical
composition of a typical WNh star, but with a large mass fraction of N
($\approx 10\%$). Overall, 
chemistry and $v_{\rm exp}$ point to a massive WN star with a ring nebula.

So far,  all of the confirmed [WN] stars are of earlier spectral subtypes. It
may be that late [WN] stars have escaped detection because of their
narrower stellar lines which can be confused with nebular emission
lines, especially at low spectral resolution.  
Moreover, also some H-rich CSs show emission lines, e.g., NGC\,6543,
and are therefore classified as of WR/Of spectral type (\cite[Smith \&
Aller 1969]{smith1969}). The stellar H$\beta$ line, which is used
for spectral classification of WN/Of stars (\cite[Crowther \& Walborn 2011]{crowther2011}), 
is often blended with the nebular emission. Thus a high spectral resolution
is also needed to tell WN type central stars from Of type central
stars.

\section{Evolutionary status}

The evolutionary channel by which [WN] stars are formed is unclear.
None of the recent late TP scenarios predicts the observed surface 
compositions of PB\,8, IC\,4663, and Abell\,48. 
It was speculated whether these objects belong 
to an alternative evolutionary sequence 
[WN] $\rightarrow$ O(He) (\cite[Miszalski et
  al.\ 2012]{miszalski2012}, \cite[Werner 2012]{werner2012}) or that
[WN] stars are O(He) stars with higher masses 
(\cite[Reindl et al.\ 2014]{reindl2014}), evolving
from RCB or sdO(He) stars, which might be merger products of non-DA WDs. 
However, the merger scenario seems to be rather unlikely due 
to the long timescales involved, which are incompatible 
with the observed low ages of the PNe of the [WN] stars. Moreover, 
PB\,8 and Abell\,48 also contain hydrogen while the assumed progenitors are
H-free.

\begin{figure}[h]
\begin{center}
 \includegraphics[width=\textwidth]{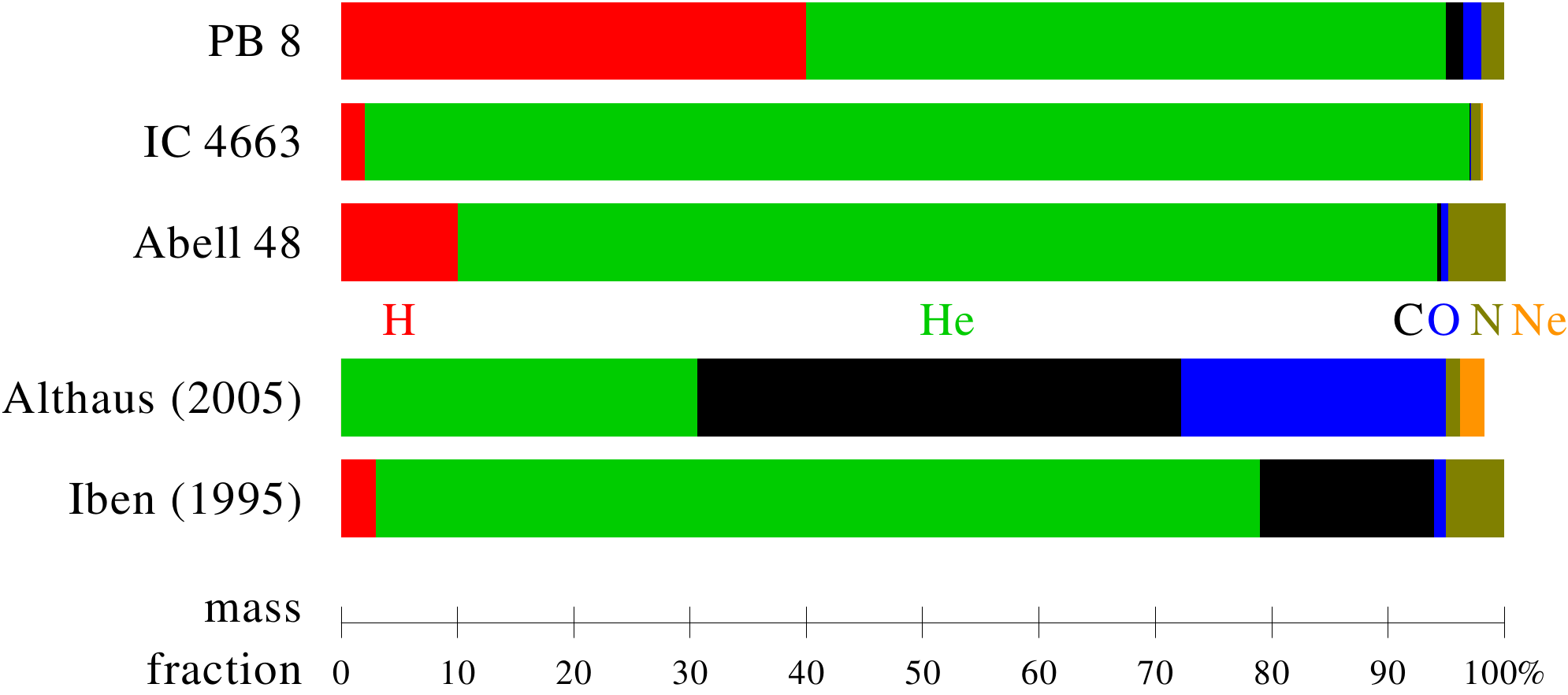} 
 \caption{Observed abundance patterns of the [WN] stars
    vs.\ predicted abundances for H-deficient central stars in
   the VLTP\,/\,born-again scenario (\cite[Althaus et al.\ 2005]{althaus2005}, \cite[Iben \&
     MacDonald 1995]{iben1995})}
   \label{fig:abundances}
\end{center}
\end{figure}

Note that the ``classical'' born-again scenario by \cite[Iben \&
  MacDonald 1995]{iben1995} yields
abundances that are somehow similar to the observed ones, including
residual hydrogen and and a supersolar nitrogen abundance (see Fig.~\ref{fig:abundances}).

\begin{thebibliography}{}

\bibitem[{{Acker} \& {Neiner}(2003)}]{acker2003}
{Acker}, A. \& {Neiner}, C. 2003, \textit{A\&A}, 403, 659

\bibitem[{{Althaus} {et~al.}(2005){Althaus}, {Serenelli}, {Panei},
  {C{\'o}rsico}, {Garc{\'{\i}}a-Berro}, \& {Sc{\'o}ccola}}]{althaus2005}
{Althaus}, L.~G., {Serenelli}, A.~M., {Panei}, J.~A., {et~al.} 2005, \textit{A\&A}, 435,
  631

\bibitem[{{Crowther} \& {Walborn}(2011)}]{crowther2011}
{Crowther}, P.~A. \& {Walborn}, N.~R. 2011, \textit{MNRAS}, 416, 1311

\bibitem[{{DePew} {et~al}\mbox{.}(2011){DePew}, {Parker}, {Miszalski}, {De
  Marco}, {Frew}, {Acker}, {Kovacevic}, \& {Sharp}}]{depew2011}
{DePew} K., {Parker} Q.~A., {Miszalski} B., {De Marco} O., {Frew} D.~J.,
  {Acker} A., {Kovacevic} A.~V., {Sharp} R.~G., 2011, \textit{MNRAS}, 414, 2812

\bibitem[{{Frew} {et~al.}(2014){Frew}, {Boji{\v c}i{\'c}}, {Parker}, {Stupar},
  {Wachter}, {DePew}, {Danehkar}, {Fitzgerald}, \& {Douchin}}]{frew2014}
{Frew}, D.~J., {Boji{\v c}i{\'c}}, I.~S., {Parker}, Q.~A., {et~al.} 2014,
  \textit{MNRAS}, 440, 1345

\bibitem[{{Garc{\'{\i}}a-Rojas} {et~al.}(2009){Garc{\'{\i}}a-Rojas},
  {Pe{\~n}a}, \& {Peimbert}}]{garcia-rojas2009}
{Garc{\'{\i}}a-Rojas}, J., {Pe{\~n}a}, M., \& {Peimbert}, A. 2009, \textit{A\&A}, 496,
  139

\bibitem[{{Herwig}(2001)}]{herwig2001}
{Herwig}, F. 2001, \textit{Ap\&SS}, 275, 15

\bibitem[{{Iben} \& {MacDonald}(1995)}]{iben1995}
{Iben}, Jr., I. \& {MacDonald}, J. 1995, in Lecture Notes in Physics, Berlin
  Springer Verlag, Vol. 443, White Dwarfs, ed. D.~{Koester} \& K.~{Werner}, 48

\bibitem[{{M{\'e}ndez}(1989)}]{mendez1989}
{M{\'e}ndez}, R.~H. 1989, in IAU Symposium, Vol. 131, Planetary Nebulae, ed.
  S.~{Torres-Peimbert}, 261--272

\bibitem[{{Miszalski} {et~al.}(2012){Miszalski}, {Crowther}, {De Marco},
  {K{\"o}ppen}, {Moffat}, {Acker}, \& {Hillwig}}]{miszalski2012}
{Miszalski}, B., {Crowther}, P.~A., {De Marco}, O., {et~al.} 2012, \textit{MNRAS}, 423,
  934

\bibitem[{{Morgan} {et~al.}(2003){Morgan}, {Parker}, \& {Cohen}}]{morgan2003}
{Morgan}, D.~H., {Parker}, Q.~A., \& {Cohen}, M. 2003, \textit{MNRAS}, 346, 719

\bibitem[{{Reindl} {et~al.}(2014){Reindl}, {Rauch}, {Werner}, {Kruk}, \&
  {Todt}}]{reindl2014}
{Reindl}, N., {Rauch}, T., {Werner}, K., {Kruk}, J.~W., \& {Todt}, H. 2014,
  \textit{A\&A}, 566, A116

\bibitem[{{Riesgo-Tirado} \& {L{\'o}pez}(2002)}]{riesgo2002}
{Riesgo-Tirado}, H. \& {L{\'o}pez}, J.~A. 2002, in RMxAC, ed. W.~J. {Henney},
  J.~{Franco}, \& M.~{Martos}, Vol.~12, 174--174

\bibitem[{{Smith} \& {Aller}(1969)}]{smith1969}
{Smith}, L.~F. \& {Aller}, L.~H. 1969, \textit{ApJ}, 157, 1245

\bibitem[{{Todt} {et~al.}(2010){Todt}, {Pe{\~n}a}, {Hamann}, \&
  {Gr{\"a}fener}}]{todt2010}
{Todt}, H., {Pe{\~n}a}, M., {Hamann}, W.-R., \& {Gr{\"a}fener}, G. 2010, \textit{A\&A},
  515, A83

\bibitem[{{Wachter} {et~al}\mbox{.}(2010){Wachter}, {Mauerhan}, {Van Dyk},
  {Hoard}, {Kafka}, \& {Morris}}]{wachtermauerhan2010}
{Wachter} S., {Mauerhan} J.~C., {Van Dyk} S.~D., {Hoard} D.~W., {Kafka} S.,
  {Morris} P.~W., 2010, \textit{AJ}, 139, 2330

\bibitem[{{Werner}(2012)}]{werner2012}
{Werner}, K. 2012, in IAU Symposium, Vol. 283, IAU Symposium, 196--203

\end{thebibliography}
\end{document}